\documentclass{icrc29}
\usepackage{graphicx,amssymb,amsmath,times}
\begin{document}
\title[Neutrino from extragalactic cosmic ray interactions in far infrared background]{Neutrino from extragalactic cosmic
 ray interactions in far infrared background}
\author[E.V. Bugaev, P.A.Klimai] {E.V. Bugaev, P.A.Klimai \\
         Institute for Nuclear Research, 60th October Anniversary
         Prospect 7a, 117312, Moscow, Russia
        }
\presenter{Presenter: E.V.Bugaev (bugaev@pcbai10.inr.ruhep.ru) ,
rus-bugaev-EV-abs1-he21-oral}

\maketitle

\begin{abstract}
Diffuse background of high energy neutrinos arising from
interactions of cosmic ray protons with far infrared radiation
background in extragalactic space is calculated. It is assumed
that cosmic ray spectrum at superhigh energies has extragalactic
origin and is proton dominated.The cosmological evolution of
extragalactic sources of cosmic ray protons as well as
infrared-luminous galaxies is taken into account in the
calculation.
\end{abstract}

\section{Introduction}
The extragalactic background of high energy neutrinos studied in
this paper arises from collisions of high energy cosmic ray (CR)
protons emitted by local sources (e.g., by active galactic nuclei)
in extragalactic space, with extragalactic background of infrared
photons. All previous studies (except the most recent ones
\cite{ref1,ref2}) took into account the interactions of CR protons
with relic photons ($T\cong 2.7K$) only. Evidently, other
components of the extragalactic radiation background also must be
taken into account in calculation of the extragalactic background
of high energy neutrinos. The interval of photon wavelengths which
is potentially most important from this point of view is
$(1-1000)\mu m$, i.e., the infrared region. At last fifteen years
infrared astronomy developed very intensively (see, e.g., reviews
\cite{ref3,ref4} ), and now the rather well-grounded calculation
of ENB from interactions of cosmic rays with extragalactic
infrared photons became possible \cite{ref1,ref2}.


For the concrete calculation of infrared background we chose the
paper of Dwek {\it et al} \cite{ref5}, where the synthetic
spectral energy distributions in the large diapason of values of
the total luminosities were derived, and the simple double
power-law form of the local luminosity function (i.e., the
present-day number density) of infrared galaxies suggested by
Soifer {\it et al} \cite{ref6}. As for the CR spectrum: we do not
know well enough the extragalactic cosmic ray spectrum and,
therefore,  we are forced to use the crucial hypothesis about
extragalactic origin of high energy CR's. Everywhere in our
calculations we  use the extragalactic model (the crossover energy
$\sim 3\times 10^{17} eV$) and normalize our theoretical CR
spectrum on experimental CR data.

Neutrino flux from interactions of high energy cosmic rays with
infrared background was recently calculated by Stanev \cite{ref1}.
Our work differs from that of \cite{ref1} in several respects. The
main difference is that in \cite{ref1} there is no non-trivial
cosmological evolution of infrared background: this evolution is
assumed to be the same as of microwave background radiation.
Besides, in \cite{ref1} there is no separating of contributions to
ENB from infrared and optical diapasons of the background while we
calculate ENB from far-infrared part of the radiation  background
only.

\section{Extragalactic CR proton spectrum at different epochs}
To obtain approximate expressions for the spectra and intensities
of CR protons in extragalactic space we use the cosmological
transport (kinetic) equation without integral term, i.e., we work
in the continuous energy loss approximation introduced, for these
problems, by Berezinsky and Grigor'eva \cite{ref7}. This equation
is written as
\begin{equation}
\label{eq1} \frac{\partial n (E,z)}{\partial
z}+\frac{\partial}{\partial E} \left[\beta(E,z)n(E,z)\right] -
\frac{3n(E,z)}{1+z}=g(E,z).
\end{equation}
Here, $n(E,z)$ is the number density of CR protons with a given
redshift $z$, the function $\beta(E,z)$ is the change of proton
energy in unit interval of $z$,
\begin{equation}
\label{eq2} \beta(E,z)=\frac{E}{1+z} - E\cdot t_p^{-1}(E,z)
\frac{dt}{dz}.
\end{equation}
The first term in r.h.s. of eq. (\ref{eq2}) takes into account
adiabatic energy losses (those due to the cosmological expansion).
The function $t_p^{-1}(E,z)$ is the cooling rate of protons via
$p\gamma\to \pi X$ and $p\gamma\to pe^{+}e^{-}$ reactions at the
cosmological epoch with redshift $z$. In our approximation, the
cooling rate consists from two parts,
\begin{equation}
\label{eq3} t_p^{-1}(E,z) = t_{p,r}^{-1}(E,z) +
t_{p,infr}^{-1}(E,z),
\end{equation}
(cooling due to interactions with relic and infrared components of
the extragalactic radiation background).

The function $g(E,z)$ in r.h.s. of the kinetic equation describes
the combined source of extragalactic cosmic rays. This source
function can be written in the form
\begin{equation}
\label{eq4} g(E,z)=\rho(z)\eta(z)f(E)\frac{dt}{dz}.
\end{equation}
Here, $\rho(z)$ is the number density of local CR sources (e.g.,
AGNs) in the proper (physical) volume, $\rho(z)=\rho_0 (1+z)^3$,
$\eta(z)$ is the activity of each local source (the integrated
number of produced particles per second), $\eta(z)=(1+z)^{m}
\eta_0\theta(z_{max}-z)$. Writing this, we assume that the
cosmological evolution of cosmic ray sources can be parametrized
by power law with the sharp cut-off at some epoch with redshift
$z_{max}$ ($m$ and $z_{max}$ are considered as parameters of a
model of the combined source). At last, the function $f(E)$ in
eq.(\ref{eq4}) describes a form of the differential energy
spectrum of the local source.

\section{Far infrared extragalactic background}
For a calculation of the extragalactic radiation background it is
convenient to use the cosmological transport equation which is
analogous to that used in the previous section. The function which
must be found is the number density of infrared photons at
different cosmological epochs, $n^{IR}(E_{\gamma},z)$. The
expression for the source function of the kinetic equation is
\begin{equation}
\label{eq5}
g^{IR}(E_{\gamma},z)=\int\frac{dL}{L}\rho(z,L)S^{IR}(E_{\gamma},L)
\frac{1}{E_{\gamma}}\cdot\frac{dt}{dz}.
\end{equation}
Here, the function $S^{IR}$ is the spectral luminosity ("spectral
energy distribution") of a luminous local source (which is in this
case an infrared-luminous galaxy with a given total infrared
luminosity $L$). The function $\rho(z,L)$ in eq.(\ref{eq5}) is the
number density of infrared-luminous galaxies with a given
luminosity $L$ in the proper (physical) volume, which is connected
with the local luminosity function $\rho(0,L)$ by the relation
\begin{equation}
\label{eq6}
\rho(z,L)=\rho\left(0,\frac{L}{(1+z)^{\gamma_l}}\right)(1+z)^{3+\gamma_{d}},
\end{equation}
where $\gamma_d$ and $\gamma_l$ are parameters determining the
cosmological evolution of the luminosity of the local source, and
evolution of the comoving density of these sources, respectively.

The resulting expression for$\;$the number density of infrared
photons in extragalactic space is (neglecting energy losses of
these photons during their travelling) is
\begin{equation}
\label{eq7} n^{IR}(E_{\gamma},z)=\int \limits_{z}^{z_{max}} dx
\left(\frac{1+z}{1+x}\right)^3 \int \frac{dL}{L} \rho(x,L)
S^{IR}\left(E_{\gamma} \frac{1+x}{1+z},L\right) \cdot
\frac{1}{E_{\gamma}} \Big| \frac{dt}{dx} \Big|.
\end{equation}

The cooling rate of CR protons in infrared radiation background is
expressed through $n^{IR}$ by the formula
\begin{equation}
t_{p,infr}^{-1}(E,z)=\frac{c}{2\gamma_p^{2}} \int
\limits_{\epsilon_{th}}^{\infty} d \epsilon_{r}
\sigma(\epsilon_{r}) f(\epsilon_{r}) \epsilon_{r}\int
\limits_{\frac{\epsilon_{th}}{2\gamma_p}}^{\infty} d\epsilon
\frac{n^{IR}(\epsilon,z)}{\epsilon^2} \label{eq8}
\end{equation}
($\gamma_p=E/m_p$). Here, $\epsilon_{r}$ is the photon energy in
the CR proton rest system, $\sigma(\epsilon_{r})$ is the
photoabsorbtion cross section, $f(\epsilon_{r})$ is the relative
proton energy loss in $p\gamma$-collision (in the observer
system).

We used the following model for the cosmological evolution of
infrared sources:
\begin{eqnarray}
\label{eq9} \rho(z,L)=\rho\left(0,
\frac{L}{(1+z)^{\gamma_l}}\right) \cdot
 (1+z)^{3+\gamma_d},
 \;\;\;\;\;\;\;\;\;\;\;\;\;\;\;\;\;\;\;\;\;\;\;\;\;\;\;\;\;\;\;
\;\;\;\;\;  z\le z_{flat}, \nonumber \\
\rho(z,L)=\rho\left(0,\frac{L}{(1+z_{flat})^{\gamma_l}}\right)\cdot
 (1+z_{flat})^{\gamma_d} (1+z)^3, \;\;\;\;\;\; z_{flat} < z\le
 z_{cut},
\end{eqnarray}
and $\rho(z,L)=0$ for $z>z_{cut}$.  The parameters are as follows:
$\gamma_d=3, \gamma_l=2, z_{flat}=1, z_{cut}=4$.

\begin{figure}[!t]
\begin{center}
\includegraphics*[trim=5 12 5 12,width=0.75\textwidth]{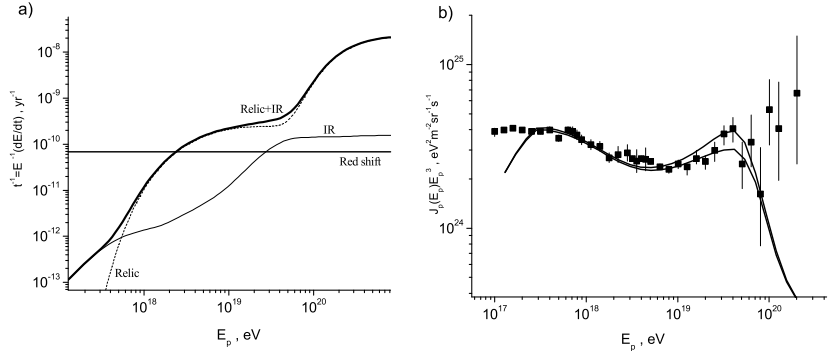}
\caption{\label {fig1} {\bf a)} Proton cooling rates in relic and
infrared photon background. {\bf b)} Extragalactic CR proton
spectrum calculated with (lower curve) and without (upper curve)
taking into account energy losses in infrared background.
  }
\end{center}
\end{figure}

On Fig. \ref{fig1}a we show the proton cooling rate on infrared
photon gas (together with corresponding function for the relic
photon case). One can see that at proton energy $\sim (3\div
4)\times 10^{19}$ eV, the contribution of infrared photons in
total cooling rate is noticeable. The results of two calculations
of extragalactic CR proton spectrum (with and without taking into
account energy losses on infrared photons) are shown on Fig.
\ref{fig1}b. The theoretical CR spectrum at $z=0$ was normalized
on experimental data; from such normalization we obtained the
value of the product $\rho_0 \eta_0$ entering eq.(\ref{eq4}) for
the source function of the CR proton kinetic equation, $\rho_0
\eta_0 \approx 1.4\times 10^{-42} cm^{-3} s^{-1} $. For the
calculation of the CR spectrum we used the following set of
parameters determining the spectrum slope and cosmological
evolution of CR sources: $\gamma=2.5, m=3.5, z_{max}=5$. We
assumed that extragalactic spectrum of CRs dominates beginning
from $E_0=3\times 10^{17}$eV.

The choice of cosmological evolution parameters $\gamma_d,
\gamma_l, z_{flat}, z_{cut}$ used in the calculation of infrared
background and CR spectrum is justified by our prediction of the
spectral intensity $E_{\gamma}I(E_{\gamma})$ of infrared
background measured experimentally (Fig. \ref{fig2}a).

\section{Extragalactic spectra of high energy neutrinos}
\begin{figure}[!t]
\begin{center}
\includegraphics*[trim=5 13 5 12,width=0.75\textwidth]{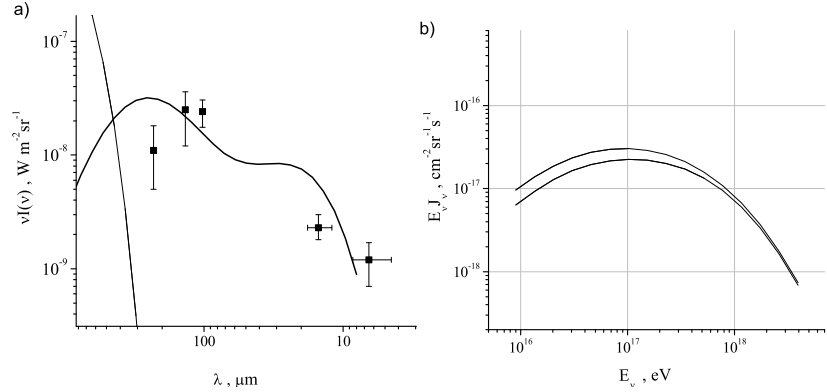}
\caption{\label {fig2} {\bf a) } Far infrared radiation background
calculated using the model of \cite{ref5} and eq.(\ref{eq9}). {\bf
b) } Extragalactic muon neutrino background: $\gamma_d=3,
\gamma_l=2, z_{flat}=1$; $z_{cut}=4$ (upper curve), $z_{cut}=3$
(lower curve).
  }
\end{center}
\end{figure}

For simplicity we use in this work one-pion approximation, i.e.,
we assume that in $p\gamma$-reaction only two particles (neutron
and pion) are produced. In this case, as is well known  the pions
have approximately isotropic distribution in the center of mass
system and, as a consequence, the step-like energy spectra in the
observer system. The main photoproduction reaction,
$p_{CR}+\gamma_{infrared}\to \pi^{+}+n$, and subsequent decays of
$\pi^+$ and $\mu^+$ lead to production of $\nu_{\mu}$,$\tilde
\nu_{\mu}$, $\nu_e$. In the present work we will be interested
only in ($\nu_{\mu}+\tilde\nu_{\mu}$)-flux. Therefore, we add
together the neutrino from pion and muon decays. Neutron produced
in the photoproduction process will decay giving proton, long
before the next interaction with photons  of the background.
Results of the neutrino spectra calculation are shown on Fig.
\ref{fig2}b.

\section{Conclusions}
In spite of the fact that the density of infrared photons in
extragalactic space is much smaller ($\sim 1.5$ photons$/$cm$^3$)
than that of relic microwave photons, the neutrino background
appears to be not so small, due to the lower threshold for
photoproduction and, especially, due to much stronger time
evolution of infrared background in comparison with that of relic
photons. The predicted neutrino fluxes are comparable, more or
less, with the neutrino fluxes from other extragalactic sources at
energy region near $10^{17}$ eV ($\gamma$-ray bursts, topological
defects, etc) and deserve further theoretical and, in future,
experimental studies.

\begin{thebibliography}{99}

\bibitem{ref1} Stanev, T., Phys. Lett. B 595, 50, 2004.
\bibitem{ref2} Bugaev, E. V., Misaki, A., and Mitsui, K., {\it astro-ph/0405109}.
\bibitem{ref3} Franceschini A., {\it astro-ph/0009121}.
\bibitem{ref4} Hauser M. G. and Dwek, E., {\it astro-ph/0105539}.
\bibitem{ref5} Dwek, E. {\it et al}, 1998, Ap. J., 508, 106.
\bibitem{ref6} Soifer, B. T. {\it et al}, 1987, Ap. J., 320, 238.
\bibitem{ref7} Berezinsky, V. S. and Grigorieva, S. I. 1988, A\&A, 199, 1.

\end{thebibliography}
\end{document}